\documentstyle[11pt,newpasp,twoside]{article}

\def\edcomment#1{\iffalse\marginpar{\raggedright\sl#1\/}\else\relax\fi}
\reversemarginpar
\input{tcilatex}

\begin{document}

\title{Why AGN Studies Need Higher Resolution}
\author{Martin J Rees}

\affil{Institute of Astronomy, Madingley Road, Cambridge CB3 OHA, UK}

\begin{abstract}
The need for high angular resolution is emphasised, especially in the 
context of programs to understand massive black holes and the processes in 
their environment.
\end{abstract}

\section{Introduction}

\subsection{Resolution limits in different wavebands}
   
 For all cosmic objects, we would ideally like to know the flux density 
over the entire electromagnetic spectrum. This goal is now almost 
achieved, at least for the brightest objects in each important class.  The 
obvious next goal is to map out each object's structure in as much detail 
as possible.   The best available angular resolution  is in the radio band: 
it is currently $10^{-3} - 10^{-4}$ arc sec from VLBI techniques, and  will further improve 
with space-based interferometry. It also, for a given baseline, improves as 
it becomes feasible to operate at shorter wavelengths.
  
   Even  sharper angular resolution in the radio band  -- better than $10^{-5}$  arc sec 
-- is achieved via the monitoring of interstellar scintillation.  Such 
scintillations offer the only probe we have of, for instance,  the emitting 
region in pulsar magnetospheres.  Another  impressive recent inference from 
interstellar  scintillation  related to  the interpretation of gamma-ray 
burst afterglows. The weak radio emission triggered by  the burst 970508 
displayed violent fluctuations during the first month, but became steadier 
thereafter (Frail et al. 1997). This implied that after one month it had become larger than the 
characteristic scale of interstellar fluctuations, 3-5 mass,  just as would 
have been expected for a  cosmologically distant object that had been 
expanding at an apparent transverse speed of $\sim 3c$  -- a rate consistent with 
blast wave models for the afterglow.

    In the visible band, adaptive optics allows the resolution  of single 
ground-based  telescopes to approach the diffraction limit, and to surpass 
that of the the Hubble Space Telescope. Optical interferometers, on the 
ground and then in space, will soon  achieve  further improvement by several 
powers of ten. Resolution in the X-ray band is now  better than 1 arc sec. There 
are already studies for X-ray interferometers, though the implementation of 
this technique remains futuristic.

\subsection{Variability and spatial resolution: complementary lines of evidence}
     It will be a long time before the most compact sources are resolved, 
but in the meantime a complementary type of information is available: 
time-variations in the flux. Variability is of course routinely  studied: 
on timescales of milliseconds in pulsars and compact stellar-mass X-ray 
sources; hours,  days and years in binary and pulsating stars, supernova 
remnants, and AGNs.

    In most instances we can study either spatial structure or variability, 
but not both: variable objects are generally too compact to be spatially 
resolved; conversely,   variability is too slow to be discernible in 
structures large enough to be mapped.  (This is exemplified, for instance, 
by the demarcation between spectroscopic binaries and visual binaries.) 
There are nearby objects within our Galaxy  which are of course 
exceptions.  But there are only two classes of extragalactic objects where 
we already have both types of data, at least in the radio band:
    
 (i) Jets with relativistic  outflow  display superluminal apparent 
velocities that  can be spatially resolved even out to the Hubble radius: 
changes in structure and flux can both be monitored.

 (ii) Slower speeds, of order the virial velocity, are of course harder to 
detect. VLBI has succeeded, however, in mapping the H$_2$O maser emission from 
moving  blobs in a disc orbiting the nucleus of the peculiar spiral galaxy 
NGC 4258.

   (As we shall hear later in the meeting,  our own Galactic Centre is 
sufficiently close that the  orbital motions  of stars   are being tracked 
in the near infrared with ever-improving precision.)

 \section{AGNs and Supermassive Black Holes}

 \subsection{The AGN population}

        The power output from active galactic nuclei (AGNs) emerges
over the entire electromagnetic spectrum, and on a range of scales
spanning almost ten orders of magnitude -- from the dimensions of the
central `engine' itself, up to the megaparsec scale of giant radio
lobes. It seems that the `engine' involves a supermassive hole, which
also generates the relativistic jets that energise strong radio
sources. The demography of these massive holes has been clarified by
studies of relatively nearby galaxies: the nuclei of most of these
display only low-level activity, but nonetheless they harbour dark
central masses.  How did these supermassive bodies form? And are they
indeed black holes with Schwarzschild/Kerr metrics, thereby offering
real prospects of testing our theories of strong-field gravity?

 \subsection{Supermassive dark objects}

      There are now two spectacularly-convincing cases of massive collapsed 
objects in nearby galaxies. The first, in  NGC 4258, has been revealed by 
amazingly precise mapping of gas motions via the 1.3 cm maser-emission line 
of H$_2$O (Miyoshi et al. 1995; Watson \& Wallin 1994).  The spectral resolution of this microwave line is high 
enough to pin down the velocities with accuracy of 1 km/sec.  The Very Long 
Baseline Array achieves an angular resolution better than  0.5 milliarc 
seconds (100 times sharper than the HST, as well as far finer spectral 
resolution!). These observations have revealed, right in NGC 4258's core, a 
disc with rotational speeds following an exact Keplerian law around a 
compact dark mass of   $3.6 \times 10^7M_{\sun}$.

   The second utterly convincing candidate lies in our own Galactic
Centre.  An unusual radio source has long been known to exist right at
the dynamical centre of our Galaxy, probably interpretable as
low-level accretion onto a massive hole (Rees 1982; Narayan, Yi \&
Mahadevan 1995). Direct evidence used to be ambiguous because
intervening gas and dust in the plane of the Milky Way prevents us
from getting a clear optical view of the central stars, as we can in,
for instance, M31. A great deal was known about gas flows, from radio
and infrared measurements, but these were hard to interpret because
gas does not move ballistically like stars, but can be influenced by
pressure gradients, stellar winds, and other non-gravitational
influences.  The situation was transformed by remarkable observations
of stars in the near infrared band, where obscuration by intervening
material is less of an obstacle. These are presented by Ekhart and by
Ghez at this meeting. The speeds scale as $r^{-1/2}$ with distance
from the centre, consistent with a hole of mass $2.5 \times 10^6 M_{\sun}$.

     The volume enclosed by the resolved orbiting material in these two 
systems -- the gas in NGC 4258, and the fast-moving stars in our Galactic 
Centre -- is small enough that   one can rule out such a high concentration 
of dark  matter in any `conventional' form (eg a dense cluster of faint 
stars).  It is important to note, however, that the scale being probed is 
still very much larger than the putative black holes themselves. The 
observed molecular disc in NGC 4258 has an orbital speed is of order 1000 
km/s. This corresponds to  radii $10^5$ times larger than the gravitational 
radius $r_g= GM/c^2$. The stars closest to our Galactic Centre likewise lie 
so far out from the putative  hole (their  speeds are less than 1 percent 
that of light)  that their orbits are essentially Newtonian.    The 
non-Newtonian domain, where the distinctive features of black holes would 
show up, has dimensions tens of thousands of times smaller.

\subsection{Kerr black holes?}

   We can infer from  the high luminosity and rapid variability of AGNs and 
compact X-ray sources that, of they are powered by gravitational energy, 
then `gravitational pits' exist,   deep and compact enough to allow several 
percent of the rest mass of infalling material to be radiated from a region 
that can vary on timescales as short as a few times $r_{g}/c$. But we still 
lack quantitative probes of the relativistic region. We believe in general 
relativity primarily because it has  been resoundingly vindicated in the 
weak field limit by high-precision observations in the Solar System, and in
the binary pulsar -- not because we yet have evidence for black holes with 
the precise Kerr metric.  What are the prospects of probing further in and 
testing the strong-field predictions of Einstein's theory?
      
Optical spectroscopy tells us a great deal about the gas in AGNs.
However, the line-emission in the visible band originates quite far
from the hole. This is because the innermost regions would be so hot
that their thermal emission emerges as more energetic quanta.  X-rays
are a far more direct probe of the relativistic region.  The
appearance of the inner disc around a hole, taking doppler and
gravitational shifts into account, along with light bending, was first
calculated by Bardeen \& Cunningham (1972) and subsequently by several
others. There is of course no hope (until X-ray
interferometry is developed) of actually `imaging' these inner discs.
However, the large frequency-shifts could reveal themselves
spectroscopically -- substantial gravitational redshifts would be
expected, as well as large doppler shifts (see, for instance, White et al. (1989)).

   Until recently, the  energy resolution and sensitivity of X-ray 
detectors was inadequate to permit spectroscopy of extragalactic objects. 
The ASCA X-ray  satellite was the first with the capability to  measure 
emission line profiles in AGNs.   It revealed a  convincing example
 (Tanaka et al. 1995) of 
a broad asymmetric emission line indicative of a relativistic disc,  and 
others should soon follow.   The angular momentum parameter in the Kerr 
metric  can in principle be constrained too,  because the emission is 
concentrated closer in, and so displays larger shifts,  if the hole is 
rapidly rotating, and there is some evidence that this must be the case in 
MCG--6-30-15 (Iwasawa et al. 1999).

      The Chandra and XMM/Newton X-ray satellites should be able to extend 
and refine these studies; they may offer enough sensitivity, in combination 
with time-resolution, to study flares, and even to follow a `hot spot' on a 
plunging orbit.

   The swing in the polarization vector of photon trajectories near a hole 
was long ago suggested (Connors, Piran \& Stark 1980) as another diagnostic; but this is still not 
feasible because X-ray polarimeters are far from capable of detecting the 
few percent polarization expected.

\subsection{Spin and precession}

The spin of a hole affects the efficiency of `classical' accretion 
processes, and  determines how much energy is in principle  extractable by 
the Blandford-Znajek  (1977) effect. Moreover, the typical amount of spin possessed 
by supermassive holes could tell us how they formed and grew.
    
Spin-up is a natural consequence of prolonged disc-mode accretion:  any 
hole that has (for instance) doubled its mass by capturing material that is 
all spinning the same way  would end up with  a/m being  at least 0.5.  A 
hole that is the outcome of a merger between two of comparable mass would 
also, generically, have a substantial spin. On the other hand, if it had 
gained its mass  from capturing many low-mass objects (holes, or even 
stars) in randomly-oriented orbits,  a/m would be small.

   Most of the literature on flows around Kerr holes assumes
axisymmetry.  This assumption is motivated not just by simplicity, but
by the expectation that Lense-Thirring precession would impose
axisymmetry close in, even if the flow further out were oblique and/or
on eccentric orbits.  Plausible-seeming arguments, dating back to the
pioneering paper by Bardeen \& Petterson (1975) suggested that the
alignment would occur, and would extend out to a larger radius if the
viscosity were low because there would be more time for Lense-Thirring
precession to act on inward-spiralling gas.  However, later studies,
especially by Pringle, Ogilvie, and their associates, have shown that
naive intuitions can go badly awry. The behaviour of the `tilt' is
much more subtle; the effective viscosity perpendicular to the disc
plane can be much larger than in the plane. In a thin disc, the
alignment effect is actually weaker when viscosity is low.  What
happens in a thick torus is a still unclear, and will have to await
3-D gas-dynamical simulations. There is now evidence for changing
orientations.

        The orientation of a hole's spin and the innermost flow
patterns could have implications for jet alignment. An important paper
by Natarajan \& Pringle (1998) shows that `forced precession' effects
due to torques on a disc can lead to swings in the rotation axis that
are surprisingly fast (i.e. on timescales very much shorter than the
timescale for changes in the hole's mass).
   
 These effects could be elucidated by images that were able to resolve 
the region, maybe 10--100 times larger than $r_g$, where the effective axis 
of symmetry tilts from that determined by the hole's spin to that 
determined by the angular momentum of  gas in the core of the host galaxy.

 \subsection{Prospects and goals}
   
In round numbers, the angular sizes corresponding to the gravitational 
radius $r_g = GM/c^2$  are:
{\obeylines{
Galactic Centre    5 $\mu$as
Giant holes in Virgo Cluster galaxies (eg M87)   5$\mu$as
Hole in 3C 273     0.1 $\mu$as}}

     The  monster holes such as the one in M87 have the same angular scale 
as the one in our Galactic Centre -- the latter is about 1000 times closer, 
but also 1000 times less massive. (Note that even the closest stellar-mass 
holes (cf section 2.6) have still smaller angular sizes.)

     Unfortunately  the  emission in the band where resolution is currently 
best -- the radio band --  comes from regions far larger than the hole 
itself. It is the optical non-thermal continuum, and the X-rays, that come 
from the region where strong-gravity effects are significant. Clues to the 
strong-gravity regime are, at least in the near-term, likely to be less 
direct (and are discussed further by Roger Blandford).

      Evidence for a hole's presence does not require such extreme 
resolution, because stars are affected by the gravitational field of a 
supermassive hole out to radii of order $10^6 r_g$. The integrated light 
from the stellar cusp is resolvable by ground-based optical telescopes as 
well as the HST, and it is measurements of this cusp that have revealed so 
many  holes in quiescent galaxies.  Improved optical resolution will 
obviously be able to probe a larger sample, as well as pinning down some of 
the current uncertainties about stellar dynamics within the cusp.

   As other speakers will discuss, there is a crude proportionality between 
the hole's mass  and that  of the central bulge or spheroid in the stellar 
distribution (which is of course the dominant  part of an elliptical 
galaxy, but only a subsidiary component of a disc system like M31 or our 
own Galaxy.)

    To understand and clarify this relationship, we need  evidence for (or 
against) black holes in small galaxies, and also in galaxies at high 
redshifts. Such information would help to test the popular scenario 
according to which galaxies form via a process of successive mergers. 
Issues include:

 (a) how much does a black hole grow (and how much electromagnetic energy 
does it radiate) in the aftermath of each merger?
and

 (b) how far up the `merger tree'  did the first massive holes form? A 
single big galaxy can be traced back to the stage when it was in dozens of 
smaller components with individual internal velocity dispersions  as low as 
20 km/sec. Did central black holes form  even in these small and weakly 
bound systems?

 \subsection{Scaling laws and `microquasars'}

     Two  galactic X-ray sources,  believed to involve black holes,  have 
double radio structures that  resemble miniature versions of the classical 
extragalactic strong radio sources.  Their jets display apparent 
superluminal motions across the sky, indicating that, like the 
extragalactic radio sources,  they contain plasma in relativistic bulk 
motion (eg Mirabel \& Rodriguez 2000).
   
 There is no reason to be surprised by these resemblances between
phenomena involving black holes with very different masses. Indeed,
the physics is exactly the same, apart from very simple scaling
laws. If we define $l = L/L_{Ed}$ and $\dot m = \dot M/ \dot
M_{crit}$, where $\dot M_{crit} = L_{Ed}/c^2$, then for a given value
of $\dot m$, the flow pattern may be essentially independent of $M$. Linear
scales and timescales, of course, are proportional to $M$, and
densities in the flow scale as $M^{-1}$.  The physics that amplifies
and tangles any magnetic field may be scale-independent, and the field
strength B scales as $M^{-1/2}$.  So the bremsstrahlung or synchrotron
cooling timescales go as $M$, implying that $t_{cool}/t_{dyn}$ is
insensitive to $M$; so also are ratios involving, for instance,
coupling of electron and ions in thermal plasma. Therefore, the
efficiencies and the value of $l$ are insensitive to $M$, and depend
only on $\dot m$.  Moreover, the form of the spectrum depends on $M$ only
rather insensitively.

    The kinds of accretion flow inferred in, for instance,  M87, giving 
rise to a compact radio and X-ray source, along with a relativistic jet, 
could operate just as well if the hole mass was a few solar masses rather 
than a few billions.  So we can actually study the same processes involved 
in AGNs in microquasars close at hand within our own galaxy. And these 
miniature sources may allow us to observe a simulacrum of the entire 
evolution of a strong extragalactic radio source, speeded up by a similar 
factor.

      For example, GRS 1915+105, one of the stellar-mass objects with
superluminal radio jets, displays quasi-periodic oscillations, at
around 60 Hz, which may be due to unstable flow patterns near the
hole (Morgan, Remillard \& Greiner 1996). The simple scaling arguments
given above imply that the AGNs which it resembles might equally well
display oscillations with the same cause. However, the periods would
be measured in days, rather than fractions of a second.

\section{Probing the Radiation Mechanism,  etc.}

\subsection{The AGN environment}

   It would of course be fascinating if we could `image' a black  hole 
sharply enough to map  the extreme-relativistic regime, detect the apparent 
distortions caused by the dramatic deflection of ray-paths by strong 
gravity, and watch the flow patterns vary.  But this will have to await 
optical or X-ray interferometers with sub-microsecond resolution. However, 
in the meantime interesting phenomena  can be probed on a range of larger 
scales.

     The broad line region (BLR) is typically at $\left(10^3 -
10^4\right)r_g$.  It varies on timescales down to the light-crossing
time, in response to changes in the ionizing continuum emitted closer
in.  The `reverberation mapping' technique has already yielded clues
to the size of the BLR and its internal dynamics, but it would
obviously help greatly if its time-varying spatial structure could be
imaged. Even more interesting would be optical imaging of the inner
parts of jets, and of the non-thermal `coronae' in which the BLR gas
is embedded.

\subsection{Brightness temperature limits}

  Self-absorption restricts the brightness temperature of  synchrotron 
emission.  This has the well-known consequence that any strong radio source 
emitting by the synchrotron process should be big enough to be resolvable 
by earth-based interferometers.  It also implies that only rather weak 
synchrotron sources would be small enough to exhibit interstellar 
scintillations.
   
Some quite strong blazars (IDVs) have, however, been found to exhibit intraday 
variability. 
(Kedziora-Chudczer et al. 1997; Dennett-Thorpe \& de Bruyn 2000).
According to standard models for  interstellar 
scintillation, these sources would seem to have a brightness temperature 
too high to be compatible with synchrotron radiation. (For a relativistic 
jet, this depends on the bulk Lorentz factor, but only via  a power  $\la 1$.)

   If an AGN displayed (for instance) intrinsic variability on timescales 
less than an hour at low radio frequencies, then the implied brightness 
temperature would be  $10^{20}K$, which would obviously imply some coherent 
mechanism. Some such process of course happens in pulsars (and is very 
poorly understood); however, few of the suggested mechanisms would scale to 
the weaker fields in AGNs. (Moreover, the high-temperature emission would 
also have to  evade attenuation by (ordinary) synchrotron self-absorption 
or by induced Compton scattering: these are  serious constraints on tenable 
models and geometry.)

   The brightness temperature implied by the IDVs seems at most a factor of 
10 higher than the synchrotron limit.  Perhaps  there is some atypical 
turbulence in the interstellar medium along the line of sight to these 
particular objects which induces scintillations even for sources whose 
angular size exceeds the usual estimate, in which case these sources would 
not pose a distinctive problem. However, if a new emission mechanism is 
indeed operating, possibilities would include:

  {\it Mild coherence in relativistic shocks}. This has been discussed especially 
by Benford (1992).

 {\it Cyclotron maser action in a kilogauss field}.  Whereas a synchrotron maser 
cannot exist except under unrealistic conditions, cyclotron masers occur 
more readily, and are operative in (for instance) Jupiter's magnetosphere.

\acknowledgements

     I am grateful to several colleagues, especially  Mitch Begelman, Roger 
Blandford, Andy Fabian and  Martin Haehnelt for discussions and 
collaboration on topics mentioned here. I thank the Royal Society for 
support.

\end{document}